\documentclass[12pt,preprint]{aastex}
\def\jcap{JCAP}
\usepackage{xcolor}

\begin{document}
\title{A Universal Relation Between Primordial Density-Potential Cross-correlation Coefficient and Spin Factor Distribution}
\author{Jun-Sung Moon$^{1}$, Jounghun Lee$^{2}$ and Juhan Kim$^{3}$}
\affil{$^1$Institute of Astronomy and Astrophysics, Academia Sinica, No. 1, Sec. 4, Roosevelt Rd., Taipei 106319, Taiwan}
\affil{$^2$Department of Physics and Astronomy, Seoul National University, Seoul 08826, Republic of Korea}
\affil{$^3$School of Physics, Korea Institute for Advanced Study, Seoul 02455, Republic of Korea}
\email{jsmoon.astro@gmail.com, cosmos.hun@gmail.com, kjhan@kias.re.kr}
\begin{abstract}
Recent studies have revealed that the key properties of visible galaxies like their optical sizes, stellar ages, star formation rates 
and morphologies are closely linked with the angular momenta of their host dark matter halos. According to the linear 
tidal torque theory, the halo angular momentum, as a conserved quantity, is directly proportional to the primordial 
spin factor, $\tau$, defined as the degree of misalignment between the principal axes of the initial density and 
potential Hessian matrices, which were found by numerical experiments to follow a Gamma distribution, fully characterized 
by its mean and variance. In this study, we heuristically develop an analytic expression for the mean and variance of $\tau$ 
in terms of the initial density-potential cross-correlation coefficient, $q$.  Analyzing a dataset from the Multiverse simulations 
performed for both of the flat $\Lambda$CDM and $w$CDM cosmologies, we prove that this analytic expression is universally 
valid in describing how the mean and variance of $\tau$ change with $q$, regardless of the smoothing scales for both of the 
cosmologies. Given the prior finding that the $\tau$-distribution can be reconstructed from the observable galaxy size distribution, 
this universal analytic expression may allow us to determine $q$ from the same observable via the mean and variance of $\tau$. 
We discuss a possibility of constraining the early universe physics from the reconstructed $q$ via our heuristic model, without 
suffering from cosmological degeneracies.
\end{abstract}
\section{Introduction}\label{sec:intro}

It has been recently disclosed by multiple numerical studies that the spin angular momenta of dark matter (DM) halos are one 
of the key players in establishing the physical properties of their visible galaxies such as their sizes, morphologies, formation 
epochs, colors, star formation rates, and supermassive blackhole masses 
\citep[e.g.,][]{ber-etal08,KL13,PR20,wal-etal22,lu-etal22,cad-etal22,rom-etal23,wan-etal25,liu-etal25,ML25a, xue-etal26}. 
According to the linear tidal torque theory (LTTT) based on the standard gravitational instability paradigm \citep{dor70,whi84}, 
the continual interactions between the protohalos and surrounding tidal fields before the turn-around moments originate the 
angular momenta of DM halos that were found by several observational works to be more or less conserved even after 
the turn-around moments \citep{LP02,LE07,mot-etal21,she-etal25,MO25}. 

In this LTTT picture, the spin angular momenta of protohalos are directly proportional to the degree of misalignments between 
the principal axes of the protohalo inertia and local tidal tensors at the protogalactic sites \citep{whi84,LP00,LP01}. 
Formulating a quantitative definition of this degree of misalignments and referring to it as {\it the primordial spin factor}, 
\citet{ML25a} showed that the links between galaxy properties and halo angular momenta are mainly established by 
the dependences of their formation epochs on the primordial spin factors. The roles played by the primordial spin factors in 
the formation and evolution of galaxies and their host halos are not limited to regulating their formation epochs but 
extended to driving the mass and radius-dependent spin transitions of galaxies \citep{ML23,ML24}, causing the secondary biases 
(i.e., assembly biases) \citep{LM24} and creating a distinct bimodality in morphological distributions~\citep{wan-etal25}. 

Recently, \citet{ML25a} directly investigated the dependences of the galaxy stellar properties on the primordial spin factors 
by analyzing data from high-resolution hydrodynamic simulations and discovered the existence of large amounts of mutual 
information between galaxy stellar properties measured at the present epochs and the primordial spin factors determined 
at the protogalactic stages.  Noting that the galaxy stellar sizes and primordial spin factors particularly share large amounts of 
mutual information, both following the Gamma distributions,  \citet{ML25b} developed an algorithm to reconstruct the 
distribution of primordial spin factors from the observable galaxy size distributions and verified the success of this algorithm 
against observation data.  

Given the previous finding that the protohalo inertia tensors are well approximated by the initial density Hessian tensors 
smoothed on the protohalo mass scales \citep{cod-etal15,mot-etal21}, the primordial spin factor was redefined as the degree of the 
misalignments between the density and potential Hessian tensors \citep{ML25b}, since the initial tidal tensors are nothing but the potential Hessians. 
Given this modified definition, it is naturally expected that the primordial spin factors should depend on the cross-correlations 
between initial density and potential fields, which has recently been highlighted as a new probe of early universe physics 
\citep[e.g.,][]{che-etal25}.
If a valid analytical model for the primordial spin factors as a function of the initial density-potential cross-correlations is 
developed, then it should be in principle useful as a new complementary diagnostic to constrain the initial conditions by 
measuring the observable galaxy stellar properties that were already found to depend on the primordial spin factors. 

The most desirable analytic model for this purpose should be the one that satisfies the following three conditions: 
first, it is expressed solely in terms of the initial density-potential cross correlations. Second, it is valid not only for the standard 
cosmology but also for the viable alternatives at which current observational data hint \citep[e.g.,][]{desi25}; 
third, its functional form is robust against the variation of the scales on which the two Hessian tensors are smoothed. 
Here, we attempt to develop such a model in a heuristic way and to validate it through numerical testing against high-resolution $N$-body simulations. 
In Section \ref{sec:analytic}, we lay out an heuristic modeling of the dependence of primordial spin factors on the 
initial density-potential correlations. In Section \ref{sec:numerical}, we present the results from a numerical testing of our heuristic 
model, revealing its robustness and universality. In Section \ref{sec:con}, we summarize the key results and discuss a prospect of our 
heuristic model as a new cosmological diagnostic. 

Throughout this paper, two different classes of the background cosmologies will be considered.  The first cosmological class 
depicts a universe where spacetime is flat (zero curvature), filled with the cosmological constant ($\Lambda$ characterized 
by the equation of state $w=-1$) and cold dark matter (CDM) as the most dominant form of energy and matter, respectively. 
To the other class belong the cosmologies where the dominant energy content of a universe is given as a scalar field DE 
characterized by $w\ne -1$, while its curvature is still zero and dominant matter content is still CDM.  
Following the convention, we refer to these two classes as the $\Lambda$CDM and $w$CDM cosmologies, and use 
$\Omega_{\Lambda}$ and $\Omega_{\rm de}$ interchangeably to denote the DE energy density parameter for the $\Lambda$CDM case.

\section{Heuristic Modeling}\label{sec:analytic}

The formal definition for the primordial spin factor was provided by \citet{ML25b} as
\begin{equation}
\label{eqn:tau}
\tau(R_{f1},R_{f2}) \equiv \sqrt{\frac{\widehat{I}^{2}_{12}+\widehat{I}^{2}_{23}+\widehat{I}^{2}_{31}}{\widehat{I}^{2}_{11}+\widehat{I}^{2}_{22}+\widehat{I}^{2}_{33}}}\, ,
\end{equation}
where $\widehat{\bf I}=(\widehat{I}_{ij})$ is the density Hessian matrix, given as $I_{ij}\equiv \partial_{i}\partial_{j}\delta$, smoothed 
on the scale of $R_{f1}$, expressed in the principal frame of the initial potential Hessian tensor, ${\bf T}=(T_{ij})$ (i.e., initial tidal tensor), 
smoothed on the scale, $R_{f2}$. It quantitatively represents the degree of misalignments between the principal axes of the initial 
density and potential hessian tensors.  Note that $\tau$ depends simultaneously on the two smoothing scales, $R_{f1}$ and $R_{f2}$. 

If there are perfect alignments or anti-alignments between the principal axes of the two Hessian tensors, then 
all of the three off-diagonal components, $\{\widehat{I}_{12},\widehat{I}_{23},\widehat{I}_{31}\}$, in Eq.(\ref{eqn:tau}) will 
vanish in the principal frame of ${\bf T}$, yielding $\tau=0$.  The Zel'dovich approximation \citep{zel70}, as the first order Lagrangian 
perturbation theory, indeed predicts that the two Hessian tensors have perfectly anti-aligned principal axes, in which case 
we expect $\tau=0$. Meanwhile, numerical experiments found imperfect but still quite strong anti-alignments between them in case of $R_{f1}=R_{f2}$, 
showing a very low but non-zero mean value of $\tau(R_{f1},R_{f2};R_{f1}=R_{f2})$ \citep{LP00,por-etal02}.  
Actually, given the stochastic nature of $\tau$, it is in fact more appropriate to deal with the mean and variance of $\tau$ rather than 
$\tau$ itself, to quantify its departure from the prediction of the Zel'dovich approximation \citep{zel70}. 

There are two factors that can significantly affect the mean and variance of $\tau$. First, the difference between $R_{f1}$ and $R_{f2}$. 
The larger this difference is, the more strongly the principal axes of the two Hessian tensors would be misaligned with one another, 
elevating the mean value of $\tau$ and enlarging its variance. Second, the cross-correlations between the initial density and 
potential fields. Even when the two Hessian tensors are smoothed on the exact same scale, if the initial density-potential 
correlations are low, then their principal axes would be strongly misaligned with one another, yielding large mean and variance values of $\tau$. 

Assuming the Gaussianity of the initial density and potential fields, the mean of $\tau$ can be theoretically evaluated as 
\begin{equation}
\label{eqn:mgam}
\bar{\tau} = \int\,\sqrt{\frac{\sum_{l=4}^{6}y^{2}_{l}}{\sum_{l=1}^{3}y^{2}_{l}}}\,p({\bf y}|{\bf z})p({\bf z})\,d^{6}{\bf y}d^{6}{\bf z}\, , 
\end{equation}
where ${\bf y}\equiv (\widehat{I}_{11},\widehat{I}_{22},\widehat{I}_{33},\widehat{I}_{12},\widehat{I}_{23},\widehat{I}_{31})$, 
$p({\bf y}|{\bf z})$ is the conditional multivariate-Gaussian distribution of ${\bf y}$ on the scale of $R_{f1}$ in the principal frame 
of ${\bf T}$ on the scale of ${\bf R}_{f2}$ whose six elements (its three eigenvalues and three Euler angles) 
are denoted by ${\bf z}$ \citep{BBKS86}, and $p({\bf z})$ is the multivariate-Gaussian distribution of ${\bf z}$.  
If the initial density and potential Hessian matrices were uncorrelated with maximally misaligned 
principal axes, then the conditional distribution would equal the unconditional one, i.e., $p({\bf y}|{\bf z})=p({\bf y})$, and thus the 
integration in the right-hand side of Eq.~(\ref{eqn:mgam}) would be reduced to 
\begin{equation}
\label{eqn:mtau}
\bar{\tau} = \int\,\left(\frac{\sum_{l=4}^{6}y^{2}_{l}}{\sum_{l=1}^{3}y^{2}_{l}}\right)^{1/2} p({\bf y})\,d^{6}{\bf y} .
\end{equation}
In this case, the functional form of $p({\bf y})$ can be straightforwardly derived since the six independent elements of the covariance 
matrix of ${\bf y}$ are all known to be expressed in terms of the initial density rms fluctuations \citep{BBKS86}. 

According to the linear perturbation theory, however, the Poisson equation correlates $\delta$ with $\Phi$ as 
$\delta \propto \nabla^{2}\Phi$, which implies that the Hessian matrices of $\delta$ and $\Phi$ should also be correlated with 
each other and thus $p({\bf y}|{\bf z})\ne p({\bf y})$. In other words, the full $12$-dimensional integration must be performed
for the analytic evaluation of $\bar{\tau}$. Instead of undertaking this heavy computational task, we adopt a rather heuristic 
approach to a functional expression of $\bar{\tau}$ in terms of the properties of the initial density and potential fields by 
making the following assumptions. 
\begin{itemize}
\item
The dependence of $\bar{\tau}$ on the initial conditions can be expressed solely in terms of the initial density-potential 
cross-correlation coefficient, $q$, defined as
\begin{equation}
\label{eqn:q}
q^{2}(R_{f1},R_{f2})\equiv \frac{\langle\delta(R_{f1})\Phi(R_{f2})\rangle}{\sqrt{\sigma^{2}_{\delta}(R_{f1})\sigma^{2}_{\Phi}(R_{f2})}} \,\,\,{\rm with}\,\, q>0 .
\end{equation}
Here, $\sigma^{2}_{\delta}(R_{f1})$ and $\sigma^{2}_{\Phi}(R_{f2})$ are the initial density and potential auto-correlations, respectively, 
while $\langle\delta(R_{f1})\Phi(R_{f2})\rangle$ is the initial density-potential cross-correlations. These three correlations can be computed as 
\begin{eqnarray}
\label{eqn:sigma_h}
\sigma^{2}_{\delta}(R_{f1}) &\equiv& \frac{1}{2\pi^{2}}\int^{\infty}_{0}k^{2}P_{\delta}(k)W^{2}(k,R_{f1})\,dk , \\
\label{eqn:sigma_t}
\sigma^{2}_{\Phi}(R_{f2}) &\equiv& \frac{1}{2\pi^{2}}\int^{\infty}_{0}k^{6}P_{\delta}(k)W^{2}(k,R_{f2})\,dk , \\
\label{eqn:sigma_ht}
\langle\delta(R_{f1})\Phi(R_{f2})\rangle &\equiv& \frac{1}{2\pi^{2}}\int^{\infty}_{0}k^{4}P_{\delta}(k)W(k,R_{f1})W(k,R_{f2})\,dk , 
\end{eqnarray}
where $P_{\delta}(k)$ is the linear density power spectrum and $W(k,R_{f})$ is a filter function with a scale radius $R_{f}$. 
This assumption is based on the fact that the $12\times 12$ covariance matrix for the conditional distribution, $p({\bf y}|{\bf z})$, 
is expressed in terms of $\sigma^{2}_{\delta}$, $\sigma^{2}_{\Phi}$ and $\langle\delta(R_{f1})\Phi(R_{f2})\rangle$ \citep{BBKS86}. Given that $\tau$, by its definition, 
does not depend on the magnitudes of the elements of ${\bf y}$ and ${\bf z}$,  the $12$-dimensional integration in Eq.(\ref{eqn:mtau}) 
will leave only the $q$-dependence to $\bar{\tau}$, while the dependences on the individual auto-correlations are all marginalized.  
\item
The decrease of $\bar{\tau}(q)$ from its maximum value with the increase of $q$ can be modeled as 
$\left(q/q_{c}\right)^{a}\exp\left(q/q_{c}-1\right)$. This assumption is based on the expectation that $\bar{\tau}$ should 
rapidly decrease toward zero as $q$ becomes closer $1$, since a stronger cross-correlation between $\delta({\bf x};R_{f1})$ and 
$\Phi({\bf x};R_{f2})$ will yield stronger alignments or anti-alignments between the principal axes of their Hessian matrices, driving 
$\bar{\tau}$ to exponentially diminish. The constant, $q_{c}$, represents a threshold value of $q$, above which $\bar{\tau}$ 
exponentially diminishes toward zero and below which $\bar{\tau}$ mildly decreases as a power-law function with index $a$. 
\item
The decrease of the variance of $\tau$, say $S_{\tau}$, from its maximum value with the increase of $q$ can be described by 
the same model as that of $\bar{\tau}$ but with different power-law index and threshold. 
This assumption is based on the empirical finding that the probability density function, $p(\tau)$, is well approximated by 
the following Gamma distribution \citep{ML25a}:
\begin{equation}
\label{eqn:gam}
p(\tau)  =  \frac{1}{\Gamma(k)\theta^{k}}\tau^{k-1}\exp\left(-\frac{\tau}{\theta}\right)\, ,
\end{equation}
where $\Gamma\equiv \int_{0}^{\infty} y^{k-1}e^{-y}\,dy$ is the Gamma function, and $(k,\theta)$ are two positive-definite 
parameters. The mean and variance of a Gamma-distributed variable equal $k\theta$ and  $k\theta^{2}$, respectively. 
If $\bar{\tau}=k\theta$ truly increases as $q$ decreases, then it can be expected that $S_{\tau}=k\theta^{2}$ should also increase 
as $q$ decreases. The specifications of $k$ and $\theta$ complete the description of the Gamma distribution, which indicates that 
$p(\tau)$ is fully constructed if its mean and variance are measured.  
\end{itemize}

\section{Numerical Results}\label{sec:numerical}

To numerically test the effectiveness of our heuristic formula presented in Section~\ref{sec:analytic}, we utilize datasets from the Multiverse 
DM only $N$-body simulations of $2048^{3}$ particles, performed on a periodic box of side length $1024\,h^{-1}\,{\rm Mpc}$ 
\citep{shi-etal17,par-etal19,hon-etal20} for four different sets of the initial conditions: two of them belonging to the $\Lambda$CDM 
cosmological class, characterized by two different $\Lambda$ density parameters, $\Omega_{\Lambda}=0.74$ and $0.64$; the 
other two belonging to the $w$CDM class, characterized by two different DE equation of state, $w=-0.5$ and $-1.5$, both of which have the 
same DE density parameter, $\Omega_{\rm de}=0.74$ (or equivalently, the matter density parameter, $\Omega_{m}=0.26$ 
since the spacetime is flat). The other key cosmological parameters shared by all of these four background cosmologies are set 
at the following values: the baryon density parameter, $\Omega_{b}=0.044$; the dimensionless Hubble parameter, $h=0.72$; the 
amplitude of initial density fluctuations, $\sigma_{8}=0.79$; the spectral index of $n_{s}=0.96$. 

Analyzing the DM particle snapshot at the earliest epoch corresponding to the redshift $z=99$, from which the Multiverse simulations 
started, we construct the initial density and potential Hessian matrices to determine $\bar{\tau}(q)$ and $S_{\tau}(q)$ for each 
cosmology. The computational steps that we take for the determination of these two quantities are summarized as follows:
\begin{itemize}
\item
Divide the simulation volume into $2048^{3}$ grids of equal sub-volumes. Apply the cloud-in-cell (CIC) method to the 
spatial distributions of DM particles at $z=99$ to construct the initial raw density contrast field, $\delta({\bf x})$, at each grid center. 
\item
Perform a fast Fourier transformation (FFT) of $\delta({\bf x})$ to obtain its Fourier counterpart, $\delta({\bf k})$, with Fourier wave vector, 
${\bf k}=(k_{l})$.
\item
Smooth $\delta({\bf k})$ with a Gaussian filter on the scale of $R_{f1}$ as 
$\delta({\bf k};R_{f1})=\delta({\bf k})\exp(-k^{2}R^{2}_{f1}/2)$ with $k\equiv \vert{\bf k}\vert$.
\item
In Fourier-space, construct the initial potential field, $\Phi({\bf k})=\delta({\bf k})/k^{2}$, and smooth it with a Gaussian 
filter on the scale of $R_{f2}$ as $\Phi({\bf k};R_{f2})=\Phi({\bf k})\exp(-k^{2}R^{2}_{f2}/2)$
\item
Construct the density and potential Hessian tensors at a given ${\bf k}$ as 
$I_{ln}({\bf k};R_{f1})=k_{l}k_{n}\delta({\bf k};R_{f1})$ and $T_{ln}({\bf k};R_{f2})=k_{l}k_{n}\Phi({\bf k},R_{f2})$, respectively.
\item 
Perform inverse FFTs of ${\bf I}({\bf k};R_{f1})$ and ${\bf T}({\bf k};R_{f2})$ to construct the initial density and potential 
Hessian tensor fields smoothed on the scales of $R_{f1}$ and $R_{f2}$, respectively. 
\item
At each grid, diagonalize ${\bf T}({\bf x};R_{f2})$ to find its three orthonormal eigenvectors. Using these eigenvectors as three 
column vectors, construct a rotation matrix, ${\bf U}$. Then, perform a similarity transformation of ${\bf I}({\bf k};R_{f1})$ 
to find an expression of the initial density Hessian tensor in the principal frame of ${\bf T}({\bf k};R_{f2})$, as
$\widehat{\bf I}({\bf x};R_{f1},R_{f2}) = {\bf U}^{-1}({\bf x};R_{f2})\cdot{\bf I}({\bf x};R_{f1})\cdot{\bf U}({\bf x};R_{f2})$ .
\item
Compute the primordial spin factor, $\tau({\bf x};R_{f1},R_{f2})$, by using the six elements of $\widehat{\bf I}$  
according to Eq.(\ref{eqn:tau}) at each grid.  
\item 
Split the whole range of $\tau$ values from the $2048^{3}$ grids into $n_{\tau}$ intervals of equal differential length, $\Delta\tau$, and then count the number of grids, 
$n_{\rm grid}$, where the primordial spin factor value falls in a given $j$-th interval, $[\tau_{j}-\Delta\tau/2,\tau_{j}+\Delta\tau/2]$, where $j$ runs from $1$ to $n_{\tau}$.  
\item 
Determine the probability density at the $j$-th bin as $p(\tau_{j};R_{f1},R_{f2})\equiv n_{\rm grid}({\tau})/(\Delta\tau\,N_{\rm grid})$, and the Poisson errors, 
$\sigma_{P}(\tau_{j})$, in $p(\tau_{j};R_{f1},R_{f2})$ as well. 
\item
Fit the numerically obtained $p(\tau_{j};R_{f1},R_{f2})$ to the Gamma distribution, $p(\tau_{j};k,\theta)$, given in Eq.(\ref{eqn:gam}), 
by adjusting $k$ and $\theta$. Determine the best-fit values of $(k,\theta)$ that 
minimizes $\chi^{2}(k,\theta)$:
\begin{equation}
\label{eqn:chi2}
\chi^{2}(k,\theta) \equiv \sum_{j=1}^{n_{\tau}}\frac{\left[p(\tau_{j};R_{f1},R_{f2})-p(\tau_{j};k,\theta)\right]^{2}}{\sigma^{2}_{P}(\tau_{j})}, 
\end{equation}
where $n_{\tau}$ is the total number of $\tau$-intervals. 
\item
Assuming a Gaussian distribution of $\chi^{2}$, i.e., $p(\chi^{2})\propto \exp(-\chi^{2}/2)$, determine separately the marginalized errors, $\sigma_{k}$ and 
$\sigma_{\theta}$, associated with the determinations of the best-fit values of $k$ and $\theta$, respectively.
\end{itemize}

\begin{figure}[ht]
\centering
\includegraphics[height=15cm,width=16cm]{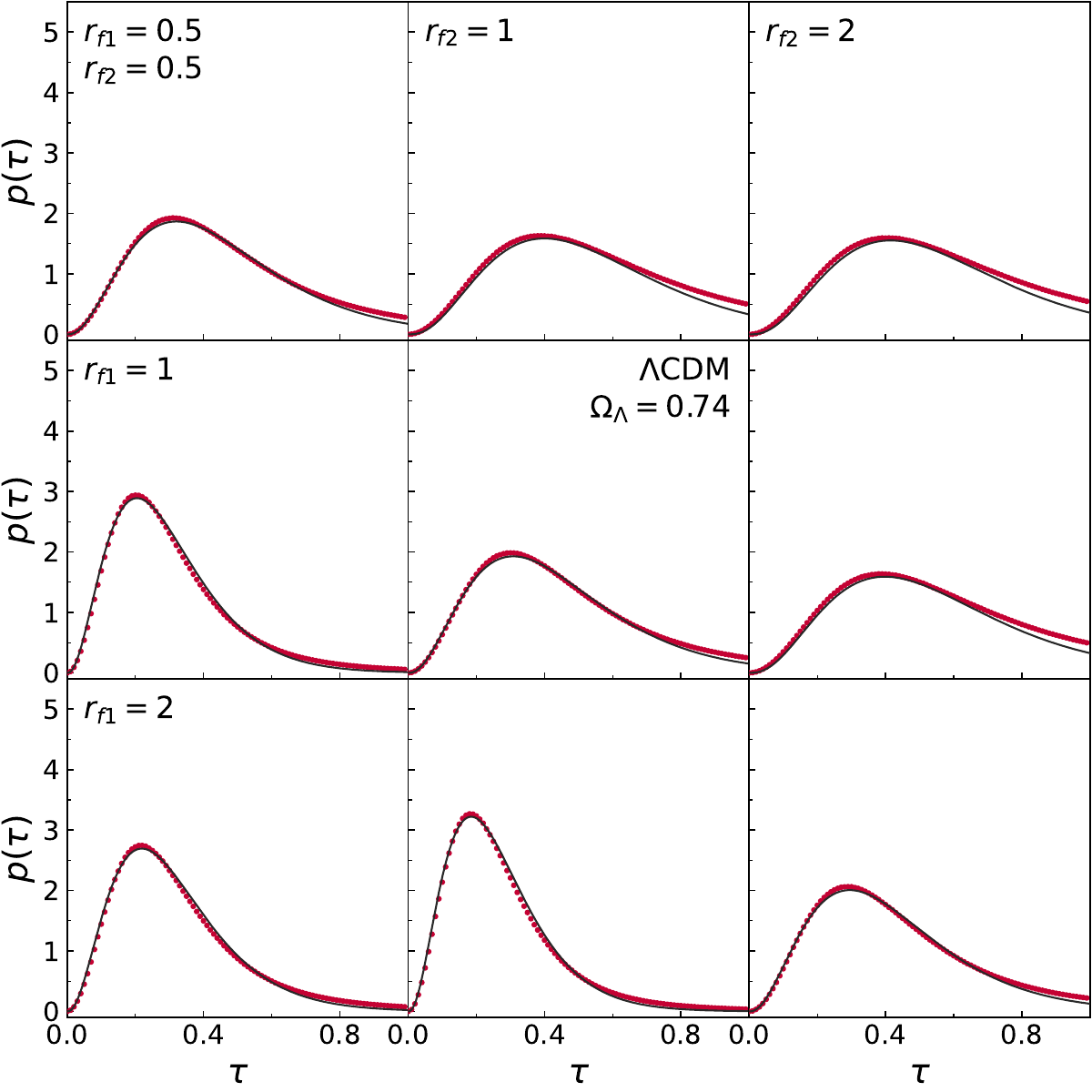}
\caption{Numerically obtained probability density function of the primordial spin factor, $\tau$, (red circles) when $\tau$ is 
obtained from the initial density and potential Hessian matrices smoothed on the scales of $r_{f1}$ and $r_{f2}$, respectively, 
for a flat $\Lambda$CDM cosmology with $\Omega_{\Lambda}=0.74$. In each panel, the numerically obtained $p(\tau)$ is 
compared with the Gamma distributions (black lines) with the best-fit parameters, $(k,\theta)$.}
\label{fig:pro_tau_ol0.74}
\end{figure}

\begin{figure}[ht]
\centering
\includegraphics[height=15cm,width=16cm]{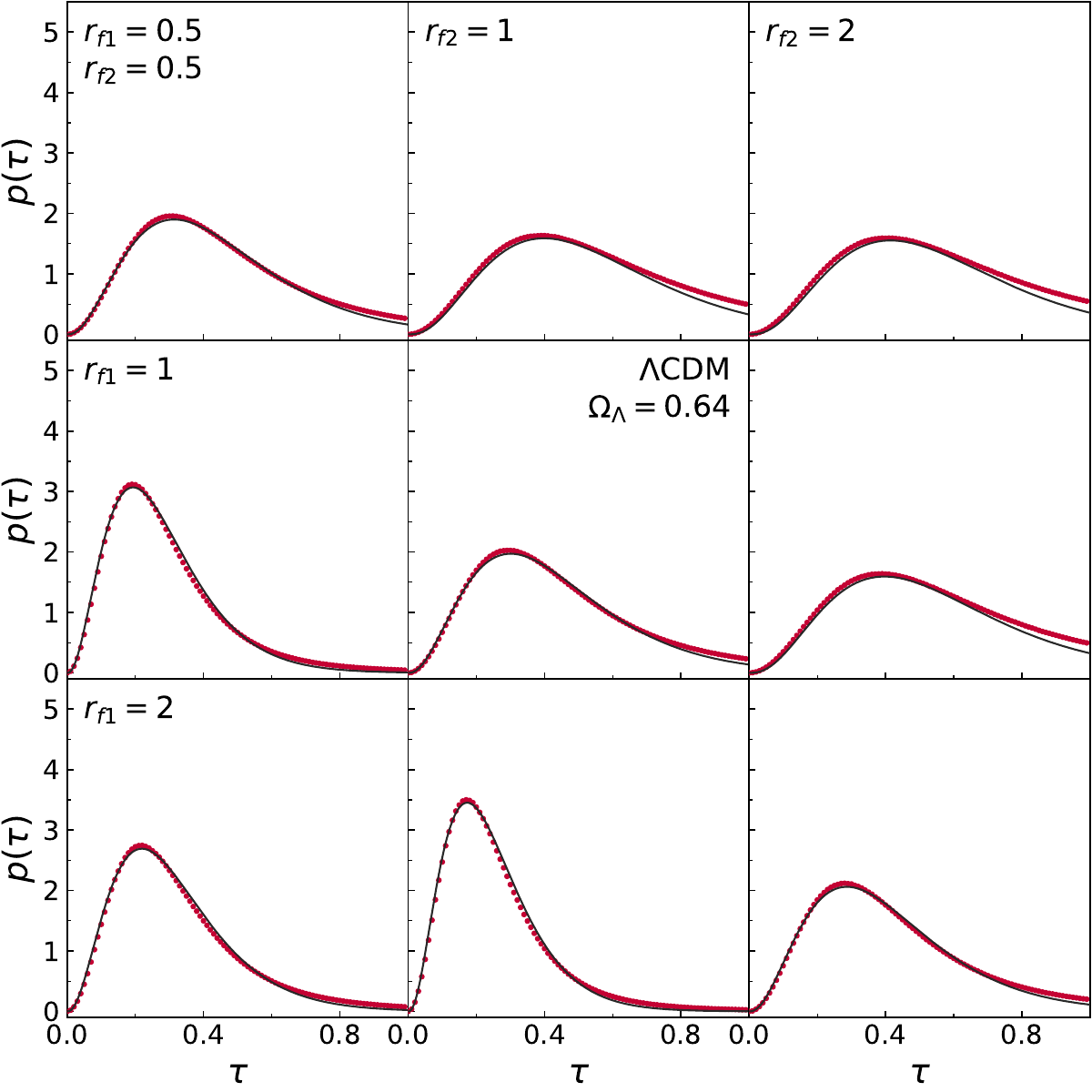}
\caption{Same as Figure \ref{fig:pro_tau_ol0.74} but for the case of $\Omega_{\Lambda}=0.64$.}
\label{fig:pro_tau_ol0.64}
\end{figure}

Figure \ref{fig:pro_tau_ol0.74} plots the numerically obtained $p(\tau)$ with Poisson errors (red circles) and compares them with the 
Gamma distribution with the best-fit parameters (black lines) for the $\Lambda$CDM cosmology with $\Omega_{\Lambda}=0.74$. 
Each of the nine panels corresponds to a different combination of $(r_{f1},r_{f2})$ with $r_{fi}=R_{fi}/(h^{-1}\,{\rm Mpc})$ for $i\in \{1,2\}$. 
Figure \ref{fig:pro_tau_ol0.64} plots the same as Figure \ref{fig:pro_tau_ol0.74} but for the case of $\Omega_{\Lambda}=0.64$.  
As can be seen, the Gamma distribution given in Eq.(\ref{eqn:gam}) agrees quite well with the numerical results, for all of the nine cases 
of $(r_{f1},r_{f2})$ for both of the cases of $\Omega_{\Lambda}$. 

It is, however, worth noting that the numerically determined $p(\tau)$ seems to have conspicuously thicker high-$\tau$ tails than the analytic model, Eq.(\ref{eqn:gam}), 
for several cases of $(r_{f1},r_{f2})$ and that the degree of this discrepancy between the numerical and analytical results seem to have a complex dependence on ($r_{f1}$, $r_{f2}$). 
At fixed $r_{f1}$, the degree of discrepancy in the high-$\tau$ section seems to increase as $r_{f2}$ increases. Whereas, at fixed $r_{f2}$, it exhibits decreasing tendency with the increment 
of  $r_{f1}$.  We suspect that this discrepancy stems from inaccuracy in the numerical determination of the principal axes of the two Hessian tensors. The high-$\tau$ values correspond 
to the cases that the principal axes of the two Hessian matrices are significantly misaligned. A spurious signal of misalignment between them, however, could be produced 
if the measurements of three principal directions of each Hessian matrix suffer from inaccuracy. There are two main factors that can lower the accuracy of the determination of the principal 
axes of the two Hessian tensors: the degree of coarseness in the field reconstruction and the degree of isotropy of a Hessian matrix. 

The first factor, i.e., the degree of coarseness, will affect mainly the determination of the principal axes of the initial density Hessian matrix. The initial raw density field 
reconstructed numerically from the particle distribution via the CIC method is rather coarse, far from being perfect. Thus, the initial density Hessian matrix 
is to inevitably  suffer from certain degree of inaccuracy in the determination of its three principal axes, which would in turn produce spurious high-$\tau$ values.
However, smoothing this initial raw-density field with a Gaussian filter on a relatively large scale will reduce its degree of coarseness and thus improve 
the accuracy of the determination of the principal axes of its Hessian matrix.  Accordingly, the probability of having a spurious signal of a high-$\tau$ value will be diminished, which 
is exactly what we have witnessed in the Figures 1-2: at fixed $r_{f2}$, the discrepancy between the numerically obtained $p(\tau)$ and the analytic Gamma distribution in the 
high-$\tau$ section decreases as $r_{f1}$ increases.

The second factor, i.e., the degree of isotropy, will affect mainly the determination of the principal directions of the initial potential Hessian matrix. Unlike the initial raw density field, the initial raw 
potential field obtained by solving the inverse Poisson equation in Fourier space is not so coarse but quite smooth, as demonstrated by the numerical work of \citet{hay-etal07}. 
When the initial raw potential field is smoothed with a Gaussian filter on a relatively large scale, its Hessian matrix becomes too isotropic, which makes it difficult to numerically determine its three 
distinct eigenvalues and directions of corresponding eigenvectors with high accuracy. In consequence, inaccurate determination of the principal axes of the initial potential Hessian matrix smoothed on 
large scales will produce spurious signals of high-$\tau$ values, which will in turn make the numerically obtained $p(\tau)$ to have thicker high-$\tau$ tails than the analytic Gamma distribution.  
This is exactly what Figures 1-2 display: at fixed $r_{f1}$, the discrepancy between the numerical and analytical $p(\tau)$ increases when $r_{f2}$ increases. 

\begin{figure}[ht]
\centering
\includegraphics[height=15cm,width=16cm]{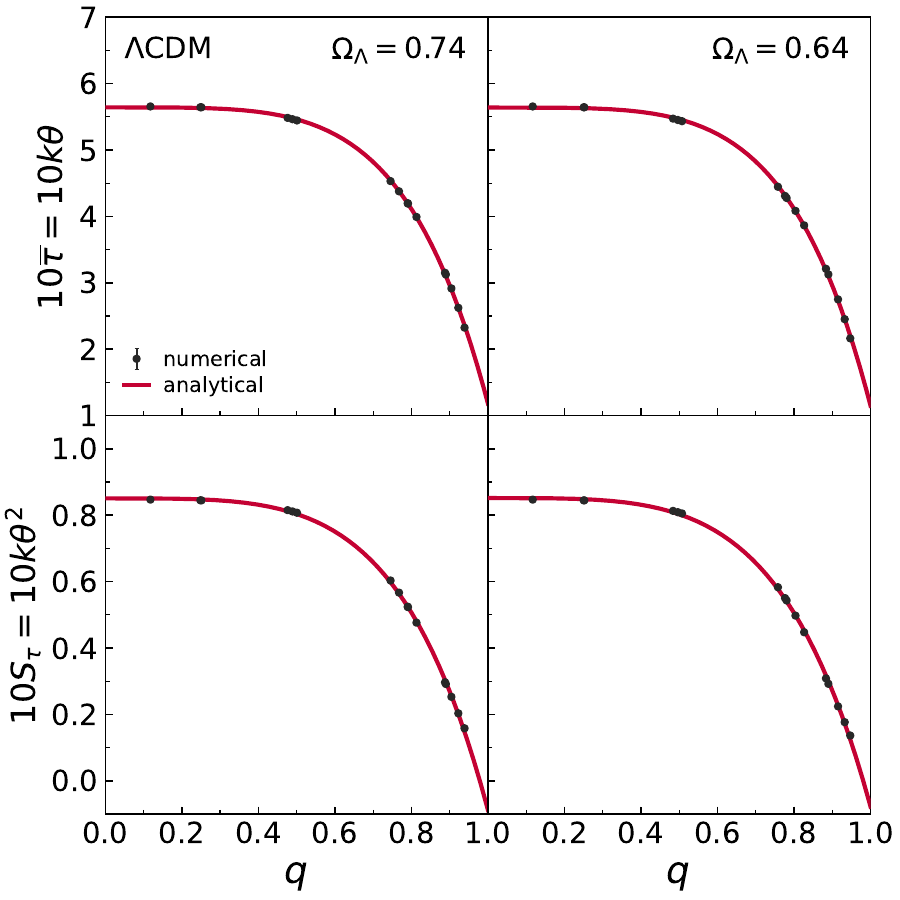}
\caption{Numerically obtained values of $\bar{\tau}=k\theta$ and $S_{\tau}=k\theta^{2}$ (black circles) versus $q$ defined in Eq.~(\ref{eqn:q}) 
compared with the fitting formula (red curves) given in Eqs.(\ref{eqn:mtau_q})-(\ref{eqn:stau_q}) for two different cases of $\Omega_{\Lambda}$.}
\label{fig:q_lcdm}
\end{figure}

Given the properties of the Gamma-distributed variables, we evaluate the mean and variance of $\tau$ as 
$\bar{\tau}=k\theta$ and $S_{\tau}=k\theta^{2}$, respectively, for various cases of $(r_{f1},r_{f2})$ including the nine cases shown in 
Figure \ref{fig:pro_tau_ol0.74}-\ref{fig:pro_tau_ol0.64}. 
As stated in Section \ref{sec:analytic},  we speculate that these mean and variance of $\tau$ should be expressed as a sole 
function of  the cross-correlation coefficient, $q$.  Computing $q$ for the same cases of $(r_{f1},r_{f2})$ by Eqs.~(\ref{eqn:q})-(\ref{eqn:sigma_ht}), 
we plot $(q,\tau)$ and $(q,S_{\tau})$ (black circles) in the top and bottom panels of Figure \ref{fig:q_lcdm}, respectively. For the computation of the 
linear density power spectrum that appears in Eqs.~(\ref{eqn:sigma_h})-(\ref{eqn:sigma_ht}), we utilize the CAMB code \citep{camb} 
with the same initial conditions as in each Multiverse simulation (corresponding to $\Omega_{\Lambda}=0.74$ and $\Omega_{\Lambda}=0.64$). 
The $q$-values for all of the cases of $(r_{f1},r_{f2})$ considered in Figures \ref{fig:pro_tau_ol0.74}-\ref{fig:pro_tau_ol0.64} are listed in Table \ref{tab:qvalue}.

As can be seen and speculated, both of $\bar{\tau}(q)$ and $S_{\tau}(q)$ exhibit exponentially diminishing behaviors in the limit of 
$q\rightarrow 1$, while mildly decreasing behavior when $q$ becomes lower than some threshold.  
We fit these numerical results of $\bar{\tau}(q)$ and $S_{\tau}(q)$ to the following formulae:
\begin{eqnarray}
\label{eqn:mtau_q}
\bar{\tau}(q) &=& \tau_{0}\left[1-\left(\frac{q}{q_{\tau,c}}\right)^{a}\exp\left(\frac{q}{q_{\tau,c}}-1\right)\right], \\
\label{eqn:stau_q}
S_{\tau}(q) &=& S_{\tau,0}\left[1-\left(\frac{q}{q_{s,c}}\right)^{b}\exp\left(\frac{q}{q_{s,c}}-1\right)\right], 
\end{eqnarray}
where $\{\tau_{0},q_{\tau,c},a\}$ and $\{S_{\tau, 0},q_{s,c},b\}$ are three constant parameters for the analytic model for 
$\bar{\tau}(q)$ and $S_{\tau}(q)$, respectively.  The top two rows of Tables \ref{tab:mtau}-\ref{tab:stau} list the best-fit values of these three parameters  
\footnote{The value of $q_{\tau,c}$ shown in Table \ref{tab:mtau} is found via the $\chi^{2}$-fitting in the extrapolated range of $q$, i.e., beyond unity.} 
for the two different cases of $\Omega_{\Lambda}$. As can be read, all of the best-fit parameters $\bar{\tau}(q)$ and $S_{\tau}(q)$ are robustly constant 
against the changes of $\Omega_{\Lambda}$ from $0.74$ to $0.64$. 
Figure \ref{fig:q_lcdm} shows how well the analytic models, Eqs.(\ref{eqn:mtau_q})-(\ref{eqn:stau_q}),  with the constant best-fit parameters (red lines) 
agree with the numerical results of $\bar{\tau}(q)$ and $S_{\tau}(q)$ (black circles) for both cases of $\Omega_{\Lambda}$. 

\begin{figure}[ht]
\centering
\includegraphics[height=15cm,width=15cm]{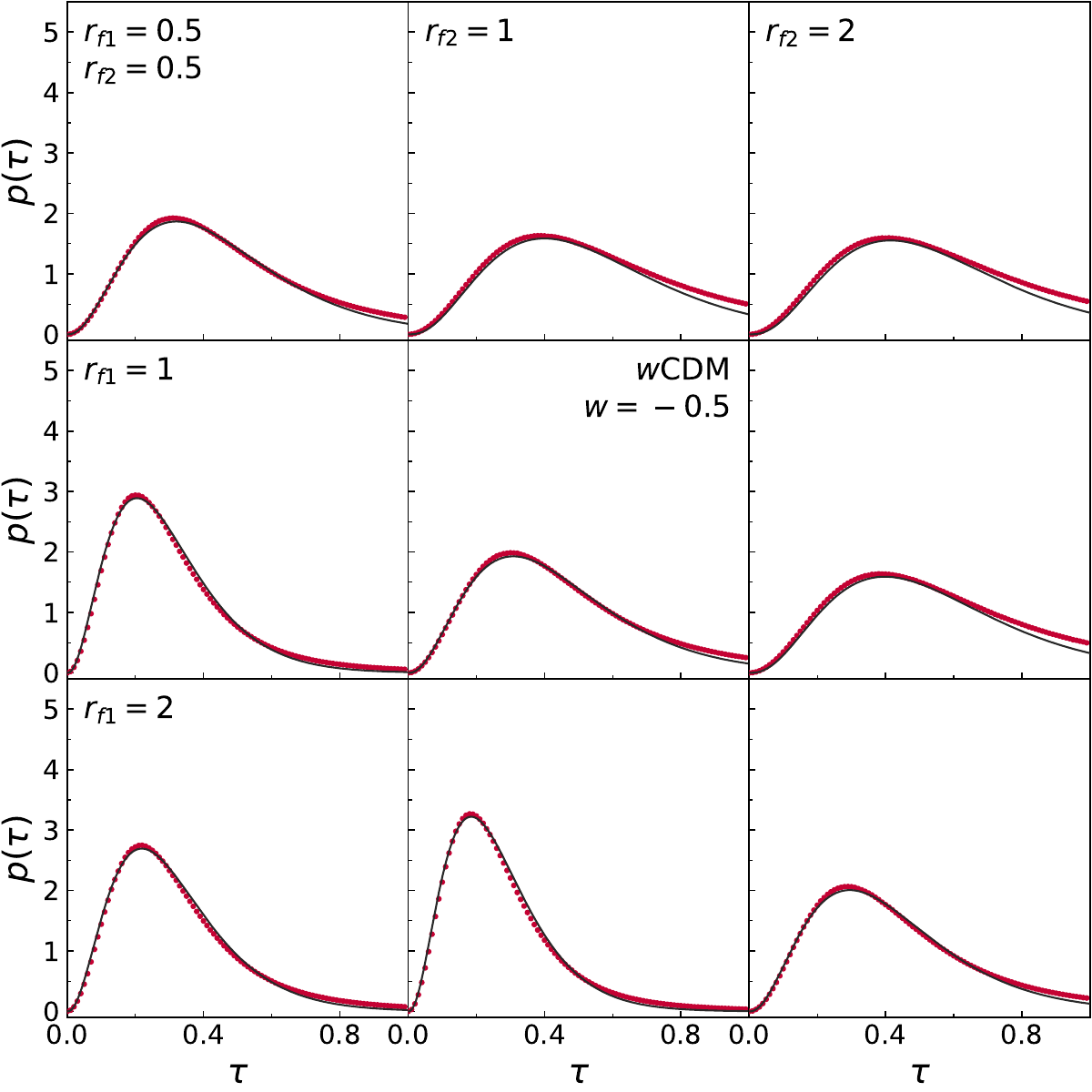}
\caption{Same as Figure \ref{fig:pro_tau_ol0.74} but for the case of a $w$CDM cosmology with $w=-0.5$.}
\label{fig:pro_tau_w0.5}
\end{figure}

\begin{figure}[ht]
\centering
\includegraphics[height=15cm,width=15cm]{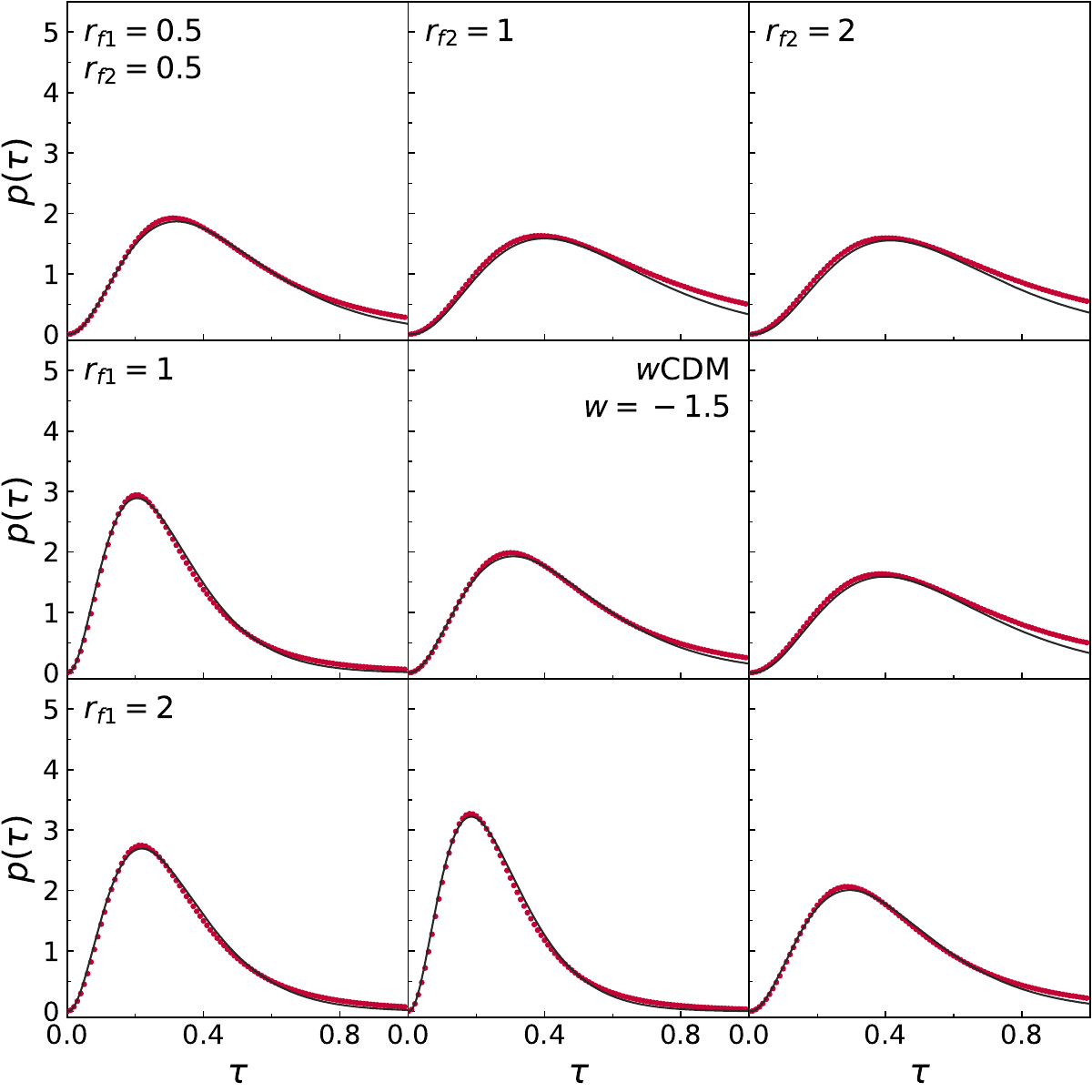}
\caption{Same as Figure \ref{fig:pro_tau_w0.5} but for the case of a $w$CDM cosmology with $w=-1.5$.}
\label{fig:pro_tau_w1.5}
\end{figure}

\begin{figure}[ht]
\centering
\includegraphics[height=15cm,width=16cm]{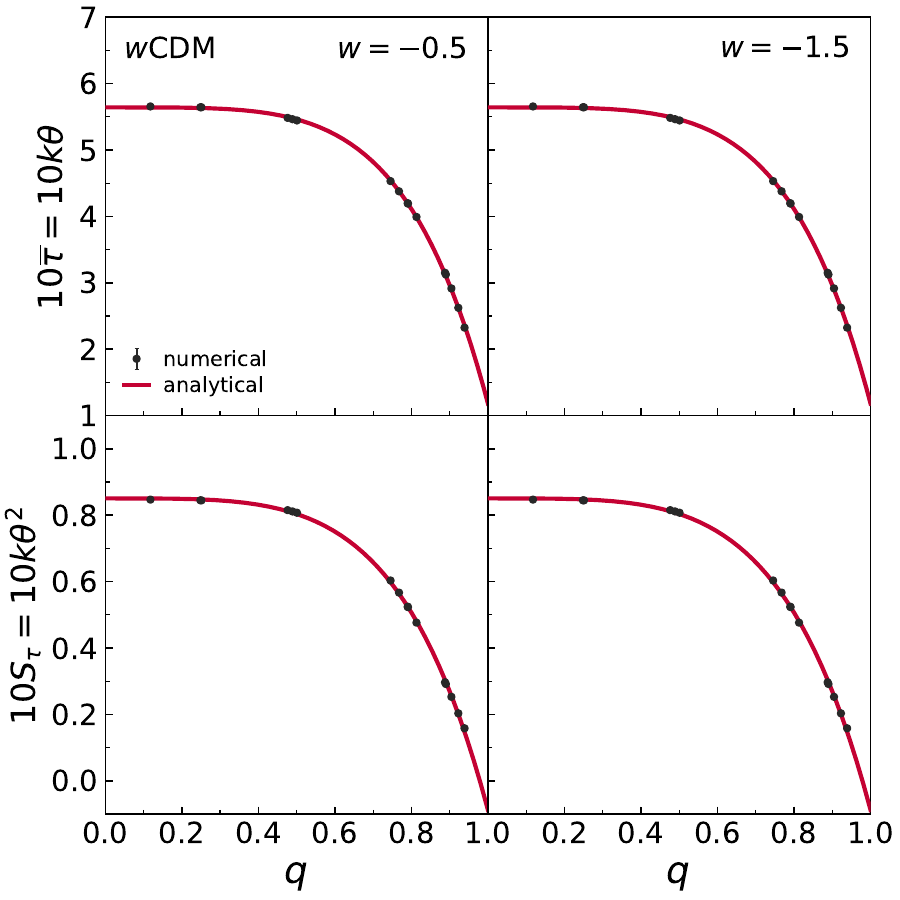}
\caption{Same as Figure \ref{fig:q_lcdm} but for the $w$CDM cases.}
\label{fig:q_wcdm}
\end{figure}

We repeat the exactly same analyses but with the datasets from the Multiverse simulations performed for the two $w$CDM 
cosmologies, the results of which are shown in Figures \ref{fig:pro_tau_w0.5}-\ref{fig:q_wcdm} and in the bottom two rows of 
Tables \ref{tab:mtau}-\ref{tab:stau}. The values of $q(r_{f1},r_{f2})$ obtained for the $w$CDM cosmologies are the same as those for the 
$\Lambda$CDM cosmologies with $\Omega_{\Lambda}=0.74$ listed in the top four rows of Table~\ref{tab:qvalue}.  As can be seen, the primordial spin factor follows well the 
Gamma distribution for all of the nine cases of $(r_{f1},r_{f2})$, and our heuristic model, Eqs.(\ref{eqn:mtau_q})-(\ref{eqn:stau_q}), excellently agree with the numerically obtained 
$\bar{\tau}(q)$ and $S_{\tau}(q)$, even for the $w$CDM cosmologies. 
Furthermore, the best-fit parameters of our heuristic model remain constant against the variation of $w$ from $-1.5$ to $-0.5$.  
These results imply that the shape and location of $p(\tau)$ may be universal over the amount and equation of state of DE.
Given that $q$ is independent of $\sigma_{8}$ and that $\bar{\tau}$ and that $S_{\tau}$ are 
expressed only in terms of $q$,  it is naturally expected that $p(\tau)$ should be also independent of $\sigma_{8}$. 

\section{Discussion and Conclusion}\label{sec:con}

We have heuristically devised an effective analytic formula for the relation between primordial density-potential cross-correlation 
coefficient and spin factor distribution, which turns out to be valid on variously different scales for both of the $\Lambda$CDM 
and $w$CDM cosmologies. We have also proven the universality of this analytic formula with the help of the Multiverse simulations 
\citep{shi-etal17,par-etal19,hon-etal20}  by showing that its three best-fit parameters remain constant against the changes of 
DE density parameter and equation of state. 
Recalling that \citet{ML25b} lately succeeded in reconstructing the primordial spin factor distribution from the observed size 
distribution of local spiral galaxies, we put forth a prospect that our analytic formula may in principle enable us to estimate the 
primordial density-potential cross-correlation coefficient from the same observables.

The initial density-potential cross-correlations provide additional information on the early universe as well as on the formation and evolution of 
large-scale structures, beyond those conveyed by the initial density and potential power spectra alone. \citet{mad-etal98} measured the cross-correlations 
between the large-scale structures and gravitational potential field to demonstrate that the large scale structures in the universe preferentially arise from the sites 
where the initial density fields are strongly cross-correlated with the gravitational potential fields.  
 \citet{gra-etal08} measured the cross-correlations between the cosmic microwave background temperature fluctuations 
and the spatial distributions of supervoids and superclusters as a proxy of the initial density-potential cross-correlations 
to detect a signal of potential decay caused by the late-time dominance of DE. 

\citet{lee14} showed that information on the initial density-potential correlations could provide a crucial key to resolving the discrepancy between the 
distant and near field cosmological probes in the value of $\sigma_{8}$ \citep{vik-etal09}. According to her theoretical calculations, the discrepancy is greatly 
alleviated if the local universe corresponds to an initial site of scale radius $100\,h^{-1}{\rm Mpc}$ where the initial density-potential cross-correlation has  
a lower value.  In the followup work, \citet{ich-etal16} proved that the same logic of \citet{lee14} would be employed to alleviate the significant Hubble tension 
\citep{htension_review}. \citet{che-etal25} theoretically verified that the cross-correlation between the initial density and potential square fields 
could be a powerful probe of the scale dependent primordial non-Gaussianity. 

In the aforementioned previous works, the initial density-potential cross-correlations were approximated by various proxies measured on the scales larger than the 
galaxy cluster scales, since the approximations are apt to fail on the smaller scales where the proxies become dominated by the nonlinear features. 
Whereas,  the heuristic model developed in the current work will make it possible to estimate directly the initial density-potential cross-correlations 
on the {\it galactic scales}. At this stage, it is worth recalling the finding of \citet{lac22} that the density three-point statistics (or equivalently bispectrum) 
on the galactic and subgalactic scales contain much larger amounts of information on the initial conditions than the linear density power spectrum on the same scales.
Recall also the claim of \citet{che-etal25} that the initial density-potential cross-correlations are as informative as the density three-point statistics on the primordial 
state of the universe although the former is much less expensive than the latter in terms of computations.  These prior findings lead us to expect that the initial 
density-potential cross-correlations, once determined on the galactic scales by our heuristic model from the galaxy properties,  
should provide new clues  to the non-trivial aspects of the early universe physics like running spectral index, scale-dependent 
primordial non-Gaussianity, formation of primordial black holes, emergence of very early dark energy and so forth \citep[e.g.,][]{KT11,bal-etal24,IA19,sob-etal25}. 

Moreover, what our heuristic model directly reconstructs from the observables is not the initial density-potential cross-correlation 
but its coefficient, $q$, which is, by definition, independent of the amplitude of the initial density power spectrum, $\sigma_{8}$. 
Given the universality of our heuristic model over $\Omega_{\rm de}$ and $w$ as well as the $\sigma_{8}$-independence of $q$, 
we also expect our diagnostic based on the heuristic model to suffer less from notorious cosmological degeneracies among the key cosmological parameters 
than the other large-scale structure diagnostics \citep{tro-etal20}.  
We plan to apply our heuristic model to real observational data to determine the initial density-potential cross-correlation coefficient 
and to explore if there is any signal of deviation from the standard physics adopted by the numerical experiments, hoping to report the result 
elsewhere in the near future. 

\acknowledgments

We thank an anonymous referee who helped us improve the original manuscript. 
JL acknowledges the support by Basic Science Research Program through the NRF of Korea funded 
by the Ministry of Education (RS-2025-00512997).

\clearpage
\clearpage
\clearpage
\clearpage
\clearpage
\clearpage
\clearpage
\begin{deluxetable}{cccccc}
\tablewidth{0pt}
\tablecaption{$\Lambda$ density parameter and initial density-potential cross-correlation coefficient, $q$, for various cases of two smoothing scales,  $r_{f1}$ and $r_{f2}$\tablenotemark{a}.}
\setlength{\tabcolsep}{3mm}
\tablehead{$\Omega_{\Lambda}$ & $r_{f1}$ & $q_{0.5}$ & $q_{1}$ & $q_{2}$ & $q_{4}$ }
\startdata
0.74 & 0.5 & 0.7454 & 0.4764 & 0.2485 & 0.1178 \\
0.74 & 1 & 0.9045 & 0.7675 & 0.4890 & 0.2511 \\
0.74 & 2&  0.8899 & 0.9226 & 0.7903 & 0.5008 \\
0.74 & 4 & 0.7909 & 0.8879 & 0.9388 & 0.8132 \\[0.5em]    
\hline\\[-0.5em]    
0.64 & 0.5 & 0.7579 & 0.4838 & 0.2504 & 0.1170 \\
0.64 & 1 & 0.9151 & 0.7806 & 0.4963 & 0.2519 \\
0.64 & 2 & 0.8898 & 0.9324 & 0.8038 & 0.5072 \\
0.64 & 4 & 0.7765 & 0.8834 & 0.9470 & 0.8265 
\enddata
\label{tab:qvalue}
\tablenotetext{a}{$q_{s}\equiv q(r_{f1},r_{f2}=s)$ for $s\in \{0.5, 1, 2, 4\}$.} 
\end{deluxetable}
\clearpage
\begin{deluxetable}{cccccc}
\tablewidth{0pt}
\tablecaption{Model, DE density parameter, DE equation of state, and three best-fit coefficients of $\bar{\tau}(q)$.}
\setlength{\tabcolsep}{3mm}
\tablehead{model & $\Omega_{\rm de}$ & $w$ & $\tau_{0}$ & $q_{\tau,c}$ & $a$ }
\startdata
$\Lambda$CDM & 0.74 & -1 & $0.564 \pm 0.001$ & $1.049 \pm 0.001$ & $3.939 \pm 0.042$ \\
$\Lambda$CDM & 0.64 & -1 & $0.564 \pm 0.001$ & $1.047 \pm 0.002$ & $3.986 \pm 0.048$ \\
$w$CDM & 0.74 & -0.5 & $0.564 \pm 0.001$ & $1.049 \pm 0.001$ & $3.939 \pm 0.042$ \\
$w$CDM & 0.74 & -1.5 & $0.564 \pm 0.001$ & $1.049 \pm 0.001$ & $3.939 \pm 0.042$ \\
\enddata
\label{tab:mtau}
\end{deluxetable}
\clearpage
\begin{deluxetable}{cccccc}
\tablewidth{0pt}
\tablecaption{Model, DE density parameter, DE equation of state, and three best-fit coefficients of $S_{\tau}(q)$.}
\setlength{\tabcolsep}{3mm}
\tablehead{model & $\Omega_{\rm de}$ & $w$ & $S_{\tau,0}$ & $q_{s,c}$ & $b$ }
\startdata
$\Lambda$CDM & 0.74 & -1 & $0.085 \pm 0.0003$ & $0.979 \pm 0.001$ & $3.583 \pm 0.062$ \\
$\Lambda$CDM & 0.64 & -1 & $0.085 \pm 0.0003$ & $0.981 \pm 0.002$ & $3.517 \pm 0.071$ \\
$w$CDM & 0.74 & -0.5 & $0.085 \pm 0.0003$ & $0.979 \pm 0.001$ & $3.583 \pm 0.062$ \\
$w$CDM & 0.74 & -1.5 & $0.085 \pm 0.0003$ & $0.979 \pm 0.001$ & $3.583 \pm 0.062$ \\
\enddata
\label{tab:stau}
\end{deluxetable}

\end{document}